\documentstyle[11pt,newpasp,twoside, epsf]{article}
\def\edcomment#1{\iffalse\marginpar{\raggedright\sl#1\/}\else\relax\fi}
\reversemarginpar

\begin{document}
\title{The Evolution of Starburst Galaxies}
 \author{Christopher J. Conselice}
\affil{Space Telescope Science Institute, USA \\
University of Wisconsin-Madison, USA}

\begin{abstract}

I review the properties of starburst galaxies in the nearby and distant
universe to decipher their evolution 
as a distinct extragalactic class.  The physical processes and environments
of massive star-formation appear to be similar out to $z\sim 4$, although the 
modes of triggering are likely quite different, varied, and still evolving.  
This is argued with the use of a structural system that measures the 
physical conditions of galaxies.  This system provides evidence that
starbursts at high-$z$ are triggered by merging, while nearby 
starbursts have a host of different triggering mechanisms, none of which, 
besides merging, are currently known to exist at $z > 2$.

\end{abstract}

\vspace{-0.5in}

\section{Introduction}

Starburst galaxies are in a fundamental phase of galaxy evolution
that possibly all galaxies undergo, and where a 
large fraction of all stars are produced.  Currently $\sim$ 25\% of
star-formation in the nearby
universe is occurring in these galaxies [14].  The fraction of
star-formation in starbursts at high redshift is almost certainly much
greater [18].  The dominate physical processes for forming stars
via a starburst are however still poorly understood. Since starbursts have 
a higher than average star-formation rate, and will exhaust their fuel of 
gas in less than a Hubble time, these events must be relatively short lived.
There must be a method, either externally or internally, that
during the evolution of a galaxy triggers the onset of massive 
amounts of star-formation.

It seems likely that mergers and interactions are primary triggering
mechanisms, particularly at high-$z$.   Despite this, there is a lack of 
conclusive proof that mergers are occurring at $z>1$.  In this paper, I 
present evidence that major mergers are occurring in a significant fraction of 
all known distant galaxies, triggering the onset of star-formation.  
At low-$z$ the situation is more complicated,
with higher order dynamical effects triggering starbursts, such as bar 
instabilities and stochastic star-formation in spiral density waves. 

\section{Local Starbursts}

Our current understanding of starbursts, and to some extent our understanding
of  high-$z$ galaxies, is derived from studying starbursts in the local 
universe.  The term starbursts was first coined by Weedman et al. [30] based
on the very blue nuclear colors of NGC 7144.  Usually the term
refers to galaxies bright in the UV, although in this paper,
a starburst is any galaxy where star-formation is higher than average,
with M$_{gas}$/\.{M}$_{SF}$ $<$ H$_{0}$$^{-1}$.

\begin{figure}
\vspace{1in}
\plotfiddle{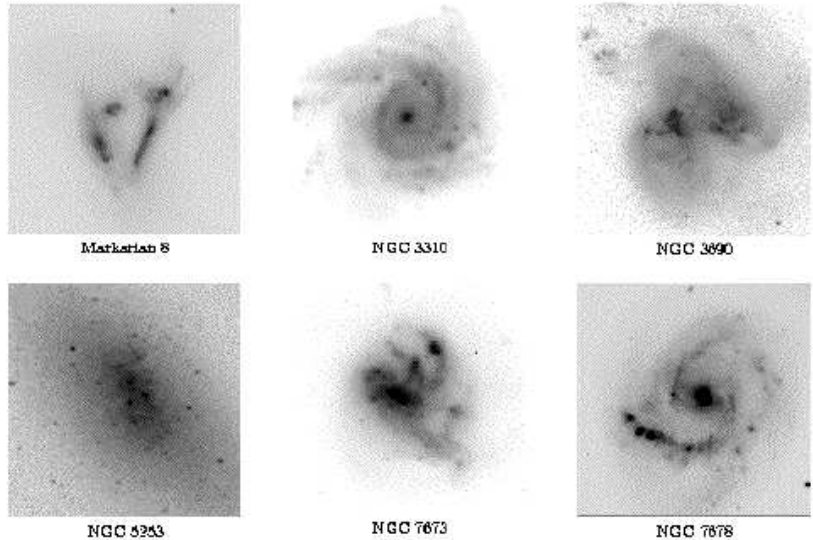}{6in}{0}{70}{70}{-200}{110}
%\plotone{config1.ps}
\vspace*{2in}
\vspace*{-6in}
\caption{Nearby UV-bright starbursts, showing the large diversity of
apparent structures.}
\end{figure}

Today we recognize several classes of nearby starburst galaxies, that I will
divide into UV-bright and dim starbursts.  The classical starbursts,
the UV-bright ones, have structures, morphologies and SEDs that are 
dominated by OB stars.  Dust extinction is occurring in these systems, but
in a patchy or `picket-fence' distribution [5,10].
The other major starburst class are those dominated by dust, and therefore
do not emit much light in the UV.  These galaxies, the (Ultra)Luminous 
Infrared Galaxies, (U)LIRGs, are undergoing massive star-formation, 
although we cannot directly view much of the light originating from the 
young OB stars due to dust attenuation.  These galaxies are re-emitting a 
large fraction of this absorbed light at far infrared wavelengths, with 
L$_{FIR} > 10^{12}$ L$_{\odot}\,$ for the ULIRGs.  

In this paper, I mainly discuss the properties of UV-bright starbursts
seen in the nearby and distant universe.  Figure 1 shows six examples of nearby
UV-bright starburst galaxies.  Anecdotally, these objects look quite
different, and one might guess their starbursts were produced
in different ways.  Markarian 8, NGC 3690 and NGC 7673 are likely merging
systems [9], while the remainder are triggered by various processes, including
an interaction with NGC 5236 (NGC 5253), an old minor merger (NGC 3310) 
and perhaps a bar instability or stochastic star-formation (NGC 7678).

Many nearby starburst are {\em not} triggered by interactions or mergers.  
Studies of paired emission line galaxies reveals that the most
luminous emission line starbursts are isolated [29].  This is not
however the case for ULIRGs, where nearly all with L$_{FIR}>10^{12}$ 
L$_{\odot}\,$ have morphological evidence of major mergers [24]. 

\section{High-$z$ Starbursts}

It is now nearly certain that many galaxies currently observed at 
redshifts $z>1$ are hosts to massive star-formation events.  This has been 
demonstrated by their SEDs (e.g., [18,27,2]) and their structural
similarities to low-$z$ starbursts (e.g., [15,10]).
The current paradigm is that UV-bright galaxies at
high-$z$ are analogs to nearby UV-bright starbursts, while galaxies
detected in the rest-frame FIR, or observed sub-mm, are dusty galaxies 
analogous to ULIRGs [21].  Other evidence for the similarity between
the gross physical processes in starbursts is the wide variety of dust 
extinctions for these objects at high and low-$z$ [2].  
High-$z$ starbursts also have a higher than average 
star-formation rate compared to nearby starbursts [31].  We know more about
high-$z$ galaxies that are bright in their rest-frame UV than at any other 
wavelength due to the relative detection sensitivities at optical 
wavelengths. We have only begun to understand galaxies where 
star-formation can be viewed at radio [22] and sub-mm wavelengths [21].  

UV-bright galaxies are also by far the most numerous high-$z$ 
galaxy known [27].  The class of galaxies detected at 850 $\mu$m are,
because of their high rest-frame FIR flux,
potentially highly enshrouded by dust [6,21].  This is further
suggested by the very faint UV counterparts to these 
galaxies [3]; a similar trend is found for nearby ULIRGs [28].  The high
850 $\mu$m background flux [13] reveals that there is possibly a large
population of these dusty star-forming galaxies that is currently only 25\%
resolved into individual objects [21].  

One huge untapped advantage for studying high-$z$ UV-bright starbursts is 
the ability to study galaxy populations based on structural 
appearances.  The structures of these galaxies (e.g., Figure 2) reveal 
important information concerning their physical states. 
If the same quantifiable structural parameters are calibrated on 
nearby starbursts [9], quiescent galaxies [7,8], and models, they 
can reveal how distant galaxies were 
triggered [9].  The results of these preliminary studies show that
many of these galaxies are mergers [11].

\begin{figure}
\vspace{-0.7in}
%\plotfiddle{config2.ps}{6in}{0}{40}{40}{-110}{110}
\plotone{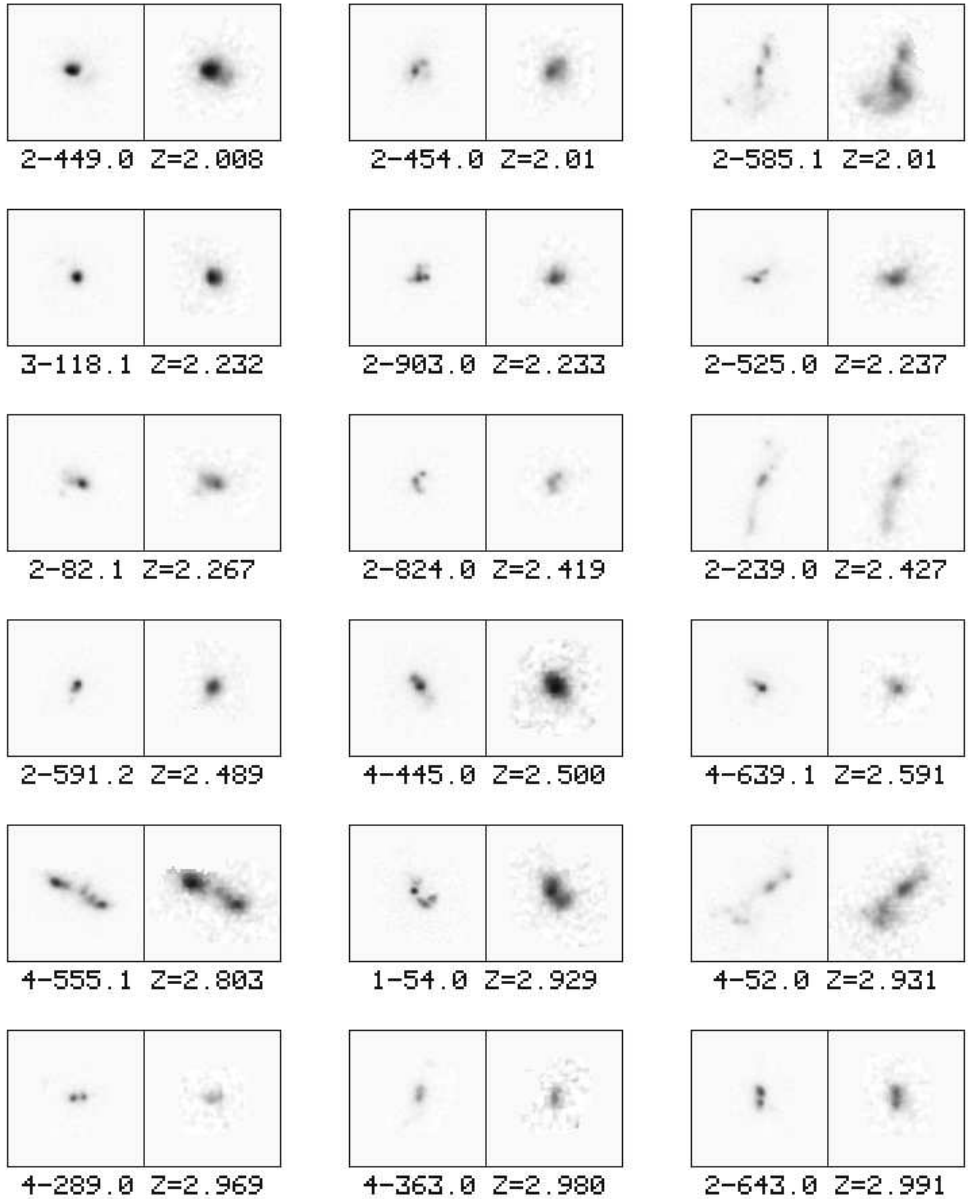}
%\vspace{-1.4in}
\caption{The rest-frame UV and optical morphologies of UV-bright $2<z<3$
star-forming galaxies located in the Hubble Deep Field [12]. 
Reproduced with the kind permission of Mark Dickinson.}
\end{figure}

\section{Dust, Metallicity, and the IMF}

Dust and metallicity effects, and perhaps non-standard initial mass
functions (IMFs), are largely responsible for the  
observational differences between galaxies undergoing
massive amounts of star-formation.    Dust is a major constitute of 
starbursts,  and it is the dominate process for altering their 
observable structures and spectral energy distributions 
[5].  Metallicity can also change the observable properties of starburst
galaxies, although current measurements indicate that high-$z$ galaxies
have metallicities similar to many local starbursts [e.g., 16].  These
measurements are however difficult and remain uncertain even for the few
galaxies studied.

\begin{figure}
\vspace{-0.7in}
\plotone{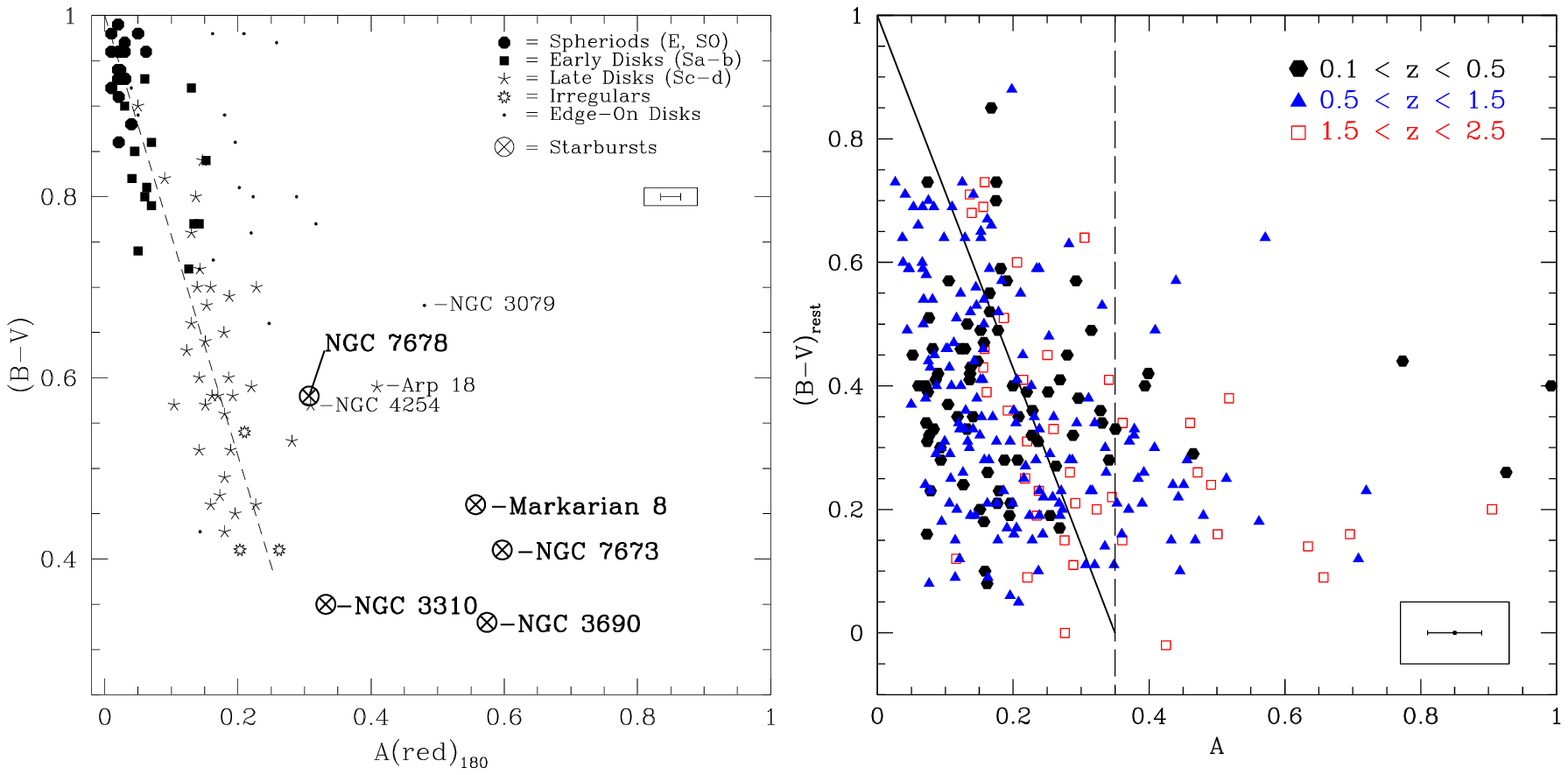}
\vspace{-4in}
\caption{(a) Color-asymmetry diagram for nearby galaxies, with mergers
located at A$>0.35$ [8,9]. (b) Color-asymmetry diagram for galaxies in
the HDF, showing the increase of major mergers at higher redshift [11].}
\end{figure}

Another question we would very much like to answer is what the IMF
of starbursts galaxies are, and if it evolves with redshift, or environment.
Derivations of the total masses and star-formation rates of starbursts rely on 
assumptions about the IMF having some form.  If a starburst's IMF is
significantly different than the current canonical
forms, derivations of most fundamental parameters are going to be
wildly off.  Unfortunately, there is no good way at present to directly
measure mass functions, besides indirect methods.  Preliminary 
observations indicated
that the IMF of starbursts are possibly truncated at lower mass limits, and
that high mass stars are overrepresented compared to solar neighborhood
Scalo and Salpeter IMFs [e.g., 23], although see [5]. Until we have a better 
understanding of the dust, metallicity and IMF properties of starbursts, 
we will be unable to accurately understand the basic physical mechanisms 
driving star-formation in these galaxies.

\section{Starburst Triggering}

One fundamental aspect of starbursts that we can study
is how they are triggered.  Starburst galaxies obviously occur after
the initial epoch of galaxy formation, since we
see them in the nearby universe.  Young stars in local starbursts are 
typically $<$ 100 Myr old [5].  This
observation tells us that galaxy evolution is still occurring, and that
existing galaxies undergo physical processes that somehow produce massive
star-formation. In the nearby universe, a starburst phase is perhaps an 
effect of evolution rather than a primary cause.  Although, at
high-$z$ the starbursting process is likely a major galaxy formation
mechanism [26].  Are methods of producing and triggering 
starbursts the same for nearby and high-$z$ galaxies?  By using the
physical techniques developed in [7,8,9,4] I will
argue that we can begin to answer this fundamental question.

The internal physical processes for producing star-formation are
still not understood, and in starbursts the process is certainly 
non-linear.  Increased gas cloud collisions
induced by larger cloud cross-sections and higher velocities may
account for these processes in interacting or merging systems, particularly
in the outer parts of galaxies. Nuclear starbursts are easier to explain due 
to the localized nature of the star-formation in compact volumes.  Barred 
galaxies also can form starbursts by funneling gas into their centers 
[19, 25]. In the nearby universe galaxies have evolved such that they 
contain these high-order bar and spiral structures, leading to a diversity 
of triggering mechanisms.  These structures are however not generally observed
at high-$z$ [1].

\begin{figure}
\plotfiddle{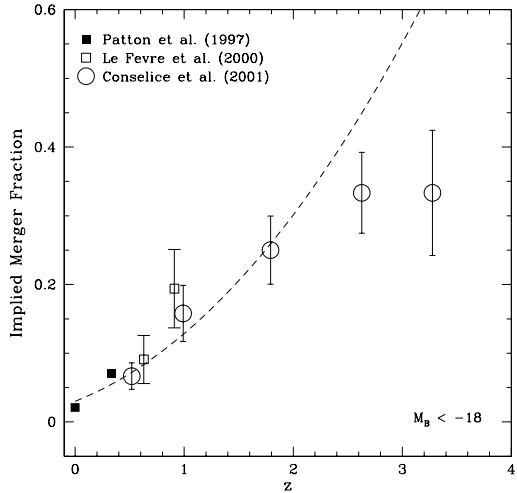}{6in}{0}{35}{35}{-110}{210}
%\plotone{con.fig2.ps}
\vspace{-3.8in}
\caption{The implied merger fraction of galaxies as a function of redshift.
Based on a combination with the pair fraction data from Le~F\'{e}vre et al. 
[17] and Patton et al. [20], our asymmetry method gives a merger fraction 
evolution with redshift f = $0.03(1+z)^{2.1\pm0.5}$.}
\end{figure}

If we want to make direct
evolutionary comparisons to high-$z$ galaxies, it is necessary to
calibrate how various parameters correlate with modes of
starburst triggering.  A basic first step is to find a method to distinguish
systems that are merging,  since it is likely a major triggering mechanism.  
This is done by using the
asymmetry parameter, a measure that indicates which galaxies are
undergoing a major merger [8].  Mergers have asymmetry values $A>0.35$ [9].
This can be seen in Figure 3a, a plot of the colors and asymmetries for
the local starbursts in Figure 1, and a sample of quiescent nearby galaxies.
The galaxies that are likely merging, based on comparisons to N-body
simulations and internal dynamics [9], have asymmetries consistent with 
merging, while the non-mergers have asymmetries $A<0.35$.  

Applying this method to high-$z$ galaxies in the 
HDF, we obtain Figure 3b. This reveals an increased merger fraction
at higher redshifts.  We can compute the merger fraction evolution with
redshift, using our criteria for identifying mergers. The
fraction of major mergers increases with redshifts as f $\sim$ 
$(1+z)^{2.1\pm0.5}$ out to z $\sim 2$ (Figure 4).  This indicates it is 
likely that a high fraction of the star-formation seen at high$-z$ is 
triggered by
the merging of similar sized galaxies.  However, only 40\% of galaxies
at $z>2$ are consistent with mergers.  The remaining starbursts
are possibly produced by minor mergers, galaxy interactions, misidentified
major mergers, or perhaps 
some other unknown triggering mechanism.  It is unlikely that these
galaxies are triggered by mechanisms related to spiral
density waves or bars, since these features are not observed at these
redshifts [1].  If these observations of the high-$z$ universe
are accurate, and if we are sampling a representative number of high-$z$
objects, then the methods of triggering massive star-formation clearly
have evolved.

I acknowledge my collaborators: J. Gallagher, M. Dickinson, M. Bershady, 
D. Calzetti, A. Jangren and N. Homeier and the NICMOS HDF-GO team 
whose active involvement in various aspects of this research have been 
invaluable. I thank Mark Dickinson for allowing me to include
Figure 2. I also thank Eva Grebel and the MPIA for financial assistance, 
and for organizing a fantastic workshop.  I appreciate support from NSF and 
NASA, especially through a Graduate Student Researchers Program (GSRP) 
Fellowship from NASA, and from the STScI Graduate Student Program.

\end{document}